\documentclass[fleqn,usenatbib]{mnras}

\usepackage[T1]{fontenc}

\DeclareRobustCommand{\VAN}[3]{#2}
\let\VANthebibliography\thebibliography
\def\thebibliography{\DeclareRobustCommand{\VAN}[3]{##3}\VANthebibliography}

\usepackage{graphicx}	
\usepackage{amsmath}	
\usepackage{amssymb}	
\usepackage{newtxtext,newtxmath}

\title[Extended Ly$\alpha$ emission around 9k quasars]{Search for extended Lyman-$\alpha$ emission around 9k quasars at z=2--3}

\author[R. Shimakawa]{
Rhythm Shimakawa$^{1}$\thanks{E-mail: rhythm.shimakawa@nao.ac.jp}\thanks{NAOJ Fellow}
\\
$^{1}$National Astronomical Observatory of Japan (NAOJ), National Institutes of Natural Sciences, Osawa, Mitaka, Tokyo 181-8588, Japan\\
}

\date{Accepted 2022 June 6. Received 2022 June 6; in original form 2022 January 16}

\pubyear{2022}

\begin{document}
\label{firstpage}
\pagerange{\pageref{firstpage}--\pageref{lastpage}}
\maketitle

\begin{abstract}
Enormous Ly$\alpha$ nebulae (ELANe) around quasars have provided unique insights into the formation of massive galaxies and their associations with super-massive black holes since their discovery.
However, their detection remains highly limited.
This paper introduces a systematic search for extended Ly$\alpha$ emission around 8683 quasars at $z=$ 2.34--3.00 using a simple but very effective broad-band $gri$ selection based on the Third Public Data Release of the Hyper Suprime-Cam Subaru Strategic Program. 
Although the broad-band selection detects only bright Ly$\alpha$ emission ($\gtrsim1\times10^{-17}$ erg~s$^{-1}$cm$^{-2}$arcsec$^{-2}$) compared with narrow-band imaging and integral field spectroscopy, we can apply this method to far more sources than such common approaches. 
We first generated continuum $g$-band images without contributions from Ly$\alpha$ emission for host and satellite galaxies using $r$- and $i$-bands. 
Then, we established Ly$\alpha$ maps by subtracting them from observed $g$-band images with Ly$\alpha$ emissions. 
Consequently, we discovered extended Ly$\alpha$ emission (with masked area $>40$ arcsec$^2$) for 7 and 32 out of 366 and 8317 quasars in the Deep and Ultra-deep (35 deg$^2$) and Wide (890 deg$^2$) layers, parts of which may be potential candidates of ELANe.
However, none of them seem to be equivalent to the largest ELANe ever found.
We detected higher fractions of quasars with large nebulae around more luminous or radio-loud quasars, supporting previous results.
Future applications to the forthcoming big data from the Vera C. Rubin Observatory will help us detect more promising candidates.
The source catalogue and obtained Ly$\alpha$ properties for all the quasar targets are accessible as online material. 
\end{abstract}

\begin{keywords}
galaxies: haloes -- galaxies: high-redshift -- intergalactic medium -- quasars: emission lines -- quasars: general 
\end{keywords}




\section{Introduction}
\label{s1}

Extended Lyman-$\alpha$ (Ly$\alpha$) nebulae around luminous quasars in the high-redshift universe provide spatial information about cool gas in the circumgalactic medium (CGM) and the intergalactic medium (IGM) around quasars.
Hence, they have a key role in understanding gas feeding and feedback mechanisms of massive galaxies and their connections to activities of super-massive black holes (e.g., \citealt{McCarthy1987,McCarthy1993,vanOjik1997,Villar-Martin2007,Faucher-Giguere2010,Goerdt2010,Tumlinson2017,Cantalupo2017,Kimock2021}, and references therein).
Multi-wavelength studies have also detected various emissions other than Ly$\alpha$ from gaseous nebulae, e.g., rest-UV helium and metal lines \citep{McCarthy1990,Maxfield2002,Villar-Martin2003,Prescott2015,Cai2017,Cantalupo2019,Guo2020,Sanderson2021}, H$\alpha$ line \citep{Shimakawa2018,Leibler2018}, dust and molecular gas (\citealt{Emonts2016,Emonts2019,Li2021}, but see \citealt{Decarli2021}).
Previous studies from a wide range of viewpoints spatially and kinetically uncovered ionisation structures by mapping multi-phase gas components on a $\sim100$ proper kpc (pkpc) scale, and addressed causal relationships between extended Ly$\alpha$ emissions and energy sources.

Highlights of recent research include discoveries of enormous Ly$\alpha$ nebulae (ELANe) around luminous radio-quiet quasars, defined as Ly$\alpha$ nebulosities with particularly high surface brightness spreading beyond hundreds pkpc \citep{Cantalupo2014,Hennawi2015,Cai2017,Cai2018,ArrigoniBattaia2018}.
Such extreme cases extending beyond the virial radius allocated an exclusive role as a probe of the underlying large-scale structures at high spatial resolution (see also \citealt{Erb2011,Fumagalli2016,Umehata2019,Kikuta2019,Daddi2021,Daddi2022}).
Discoveries of ELANe also suggested that cool gas components are more enriched over the IGM scale in and around massive haloes at high redshifts than previously considered.

Such great progress is particularly brought by the advent of integral field units (IFU), e.g., the Multi- Unit Spectroscopic Explorer on the Very Large Telescope (VLT/MUSE; \citealt{Bacon2010}) and the Keck Cosmic Web Imager on the Keck telescope (Keck/KCWI; \citealt{Morrissey2012}).
The high-performance IFUs enable intensive surveys of diffuse Ly$\alpha$ emission around 10 to 100 high-redshift quasars \citep{Borisova2016,ArrigoniBattaia2019,Cai2019,Farina2019,Drake2019,Fossati2021} to investigate the diversity of Ly$\alpha$ nebulosities depending on factors such as dynamics, luminosity, redshift, and radio loudness.
Front-line observations also show that ELANe take a crucial role as a signpost of (metal-enriched) inspiraling accretion occurring in proto-cluster haloes (\citealt{ArrigoniBattaia2018}; see also \citealt{Angles-Alcazar2017,Brennan2018,Grand2019}).
Such a new perspective provides unique insights into puzzling questions to the gas feeding mechanism in massive haloes at high redshifts \citep{Keres2005,Dekel2009,Suresh2019,Stern2020}, and star formation \citep{Chen2021,Nowotka2022,ArrigoniBattaia2021} and chemical enrichment of proto-clusters \citep{Dave2011,Shimakawa2015,Valentino2016,Vogelsberger2018,Maiolino2019}.
Thus, increasing the number of ELAN samples will help achieve consensus on wide-ranging gas feeding and feedback phenomena in high-$z$ massive haloes, but it is challenging.
\citet{ArrigoniBattaia2019} have reported that ELANe are extremely rare among luminous quasars with $-28.29\leq M_{1450}\leq-25.65$ ($\sim1$ per cent), indicating that the pruning of quasar samples is necessary to increase identifications of ELANe within reasonable observational times.

Motivated by these results, this study performs a systematic search for the extended Ly$\alpha$ emission associated with quasars at $z=$ 2--3 based on the advanced wide-field imaging data delivered by the Hyper Suprime-Cam Subaru Strategic Program on the 8.2 m Subaru Telescope (HSC-SSP; \citealt{Aihara2018,Miyazaki2018,Furusawa2018,Kawanomoto2018,Komiyama2018}). 
The HSC-SSP searches in the $grizy$ broad-bands approximately 3--4 mag deeper and with better-seeing size compared to the Sloan Digital Sky Survey (SDSS, \citealt{Gunn1998,Gunn2006,Doi2010}) and the Panoramic Survey Telescope and Rapid Response System (Pan-STARRS, \citealt{Chambers2016}) but over $>10$ times smaller survey area, e.g., $5\sigma$ limiting magnitude of $\gtrsim26$ mag and seeing full-width-half-maximum (FWHM) of $\sim0.6$ arcsec in the $i$-band (see \citealt{Aihara2022} for details). 
Various studies have demonstrated the utility of the high-quality data from the HSC-SSP, such as high-redshift galaxy and quasar surveys \citep{Ono2018,Matsuoka2016}, cluster and void search \citep{Oguri2018,Shimakawa2021}, and machine-learning-based explorations of peculiar objects \citep{Kojima2020,Tanaka2022}.
This paper adds another chapter to such past achievements in the HSC-SSP.
We aim to detect extended Ly$\alpha$ emission around more quasars than ever before by using broad-band data sets taken over a 1000 square degree field.
\citet{Prescott2012,Prescott2013} performed a similar approach to the 9.4 deg$^2$ Bo\"otes field and confirmed that it worked successfully.
It allows us to prune quasars and, hence, select prospective candidates of ELANe before more expensive narrow-band and IFU observations.
Moreover, we can investigate statistical trends of spatial properties of Ly$\alpha$ emissions in response to various physical parameters with unrivalled data sets, down to a sensitivity limit $\sim10^{-17}$ erg~s$^{-1}$cm$^{-2}$arcsec$^{-2}$, providing important insights into the relationship between super-massive black holes and the extent of Ly$\alpha$ emission.

This work is based on the public data from the Third Public Data Release of the Hyper Suprime-Cam Subaru Strategic Program (HSC-SSP PDR3; \citealt{Aihara2022}), covering approximately 1200--1300 deg$^2$ out of the entire survey footprint of 1400 deg$^2$ in $grizy$ bands.
Additionally, we adopt the final Sloan Digital Sky Survey IV (SDSS-IV; \citealt{Blanton2017}) quasar catalogue (DR16Q; \citealt{Lyke2020}) from Data Release 16 of the extended Baryon Oscillation Spectroscopic Survey (eBOSS; \citealt{Dawson2016}).
These two large databases enable an intensive search for extended Ly$\alpha$ emission around more than 9000 quasars at $z=$ 2--3 (section~\ref{s2}).
Combined with the public spec-$z$ sources at the similar redshift, we construct Ly$\alpha$ images of the targets using the redshift-corrected broad-band $gri$ colours (section~\ref{s3}) and investigate the extent of Ly$\alpha$ emission around the quasar samples (section~\ref{s4}).
We then discuss general trends of obtained Ly$\alpha$ properties against luminosity, redshifts, and radio loudness of quasars (section~\ref{s5}).
The results and conclusions achieved by this work are organised by the last section~\ref{s6}.

This research adopts the AB magnitude system \citep{Oke1983}. 
Moreover, we assume cosmological parameters of $\Omega_M=0.310$, $\Omega_\Lambda=0.689$, and $H_0=67.7$ km~s$^{-1}$Mpc$^{-1}$ in a flat Lambda cold dark matter model, which are consistent with those from the Planck 2018 VI results \citep{PlanckCollaboration2020}.


\section{SDSS quasars in HSC-SSP PDR3}
\label{s2}

This paper employs 8683 quasars at $z=$ 2.34--3.00 from the DR16Q catalogue \citep{Lyke2020} covered by the HSC-SSP PDR3 field \citep{Aihara2022}.
This section explains how and why we select these quasars as our targets.

\begin{figure}
\centering
\includegraphics[width=7.5cm]{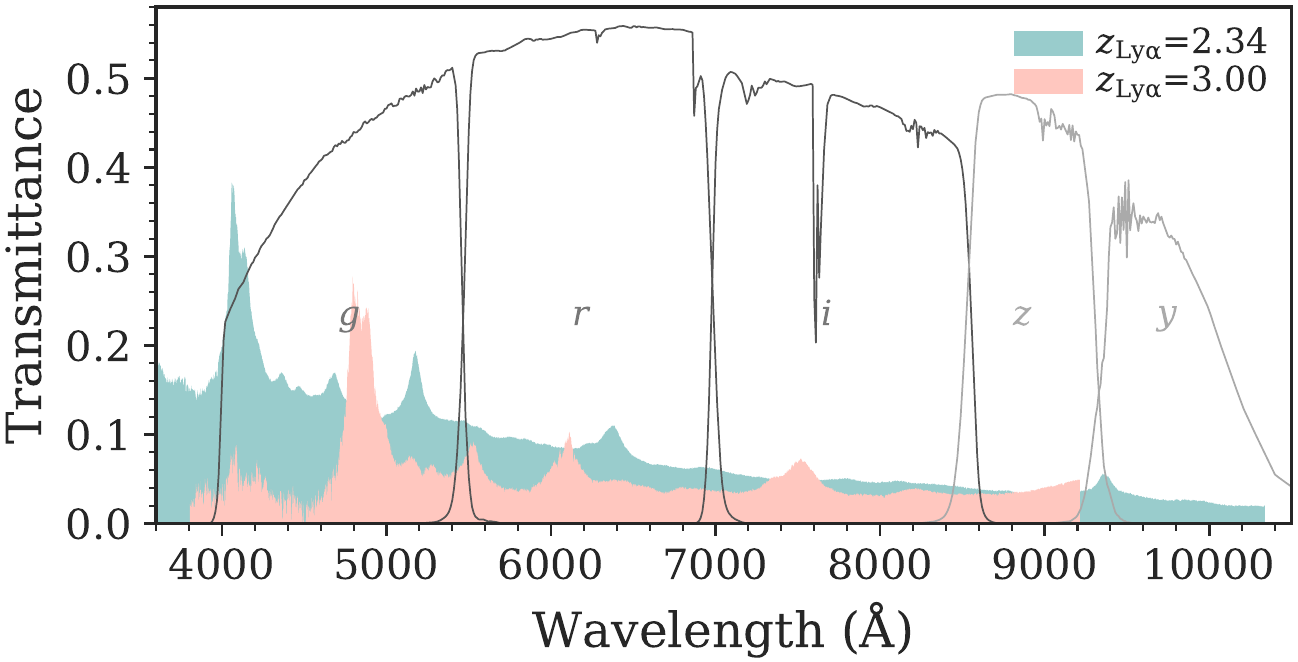}
\caption{
Effective throughput of HSC broad-band filters ($grizy$). 
The cyan and pink filled regions depict example spectra of our target quasars at $z=2.34$ and $z=3.00$, respectively (arbitrary scale).
}
\label{fig1}
\end{figure}

We began with 119,219 quasars at the Ly$\alpha$ redshift ({\tt Z\_LYA}) = 2.34--3.00 with $|\Delta z|\equiv|\mathtt{Z\_PIPE-Z\_LYA}|<0.15$ and {\tt ZWARN\_LYA = 0} in the DR16Q catalogue. 
They are originally selected based on three data sets of SDSS-IV/eBOSS \citep{Myers2015}, the Wide-field Infrared Survey Explorer \citep{Wright2010,Lang2016}, and the Palomar Transient Factory \citep{Law2009,Rau2009} over 14,000 deg$^2$, with a magnitude limit of $g<22$ or $r<22$.
Then, they are classified as quasars if their spectra taken by the eBOSS spectrographs \citep{Smee2013} are best matched to quasar models through the BOSS {\tt spec1d} pipeline (\citealt{Bolton2012}; see section~2 and 3 in \citealt{Lyke2020} for details). 
The reason why we select quasars at $z=$ 2.34--3.00 is that their Ly$\alpha$ lines fall into the $g$-band of Subaru/HSC (figure~\ref{fig1}), which enables us to extract those extended features around quasars using $gri$ broad-band colour selection (section~\ref{s3}).
In fact, the $g$-band filter can capture Ly$\alpha$ line up to $z\sim3.5$; however, we do not use quasars at $z>3$ because of heavy IGM absorption at $\lesssim1020$ \AA\ penetrating the $g$-band.
We will discuss this point in section~\ref{s3}.

\begin{table}
\caption{
SDSS/eBOSS quasars adopted in this study.
Here 290 duplicates are removed from the Wide layer sample.
}
\label{tab1}
\begin{tabular}{lrrr} 
	\hline
	Layer & Area (deg$^2$) & \multicolumn{1}{l}{Depth (min-max) [$g,r,i$]} & \multicolumn{1}{l}{N}\\
	\hline
	DUD   & 35             & [26.0-28.5, 25.7-28.1, 25.3-28.1]   & 366\\
	Wide  & 890            & [25.8-27.1, 25.4-27.0, 25.2-26.8]   & 8317\\
	\hline
\end{tabular}
\end{table}

After that, we cross-matched the quasar samples with the HSC-SSP PDR3 sources within a radius of 1 arcsec.
The HSC-SSP PDR3 field consists of the Deep and Ultra-deep (DUD) and Wide layers, respectively, covering 37 and 1332 deg$^2$ in 278 nights of observation \citep{Aihara2022}.
The HSC-SSP PDR3 gives us a science-ready catalogue and coadd data, which were well reduced by the dedicated pipeline ({\tt hscPipe version 8}; \citealt{Bosch2018}). 
The survey area is split on the database into approximately $1.7\times1.7$ deg$^2$ areas, termed {\tt tracts}, and further divided into approximately $12\times12$ arcmin$^2$ regions called {\tt patches} \citep{Aihara2022}. 
To remove the data taken under poor sky conditions, we discarded patches with seeing FWHM $>1.0$ arcsec or relatively shallow imaging depths in either of the $gri$ bands (see the limiting magnitude range in table~\ref{tab1}). 
This quality management reduced the survey area to 35 and 890 deg$^2$ in the DUD and Wide layers.
Furthermore, we selected quasars not affected by nearby bright stars ($G<18$ mag; \citealt{Aihara2022}) and bad pixels in the $grizy$-band data of the HSC-SSP PDR3, by applying the following criteria in the {\tt SQL} query:
\begin{description}
\item[--]{\tt isprimary=True},
\item[--]{\tt inputcount\_flag\_noinputs=False},
\item[--]{\tt pixelflags\_edge=False},\
\item[--]{\tt pixelflags\_bad=False},
\item[--]{\tt mask\_brightstar\_halo=False},
\item[--]{\tt mask\_brightstar\_ghost=False},
\item[--]{\tt mask\_brightstar\_blooming=False}.
\end{description}
For details on these catalogue flags, refer to \citet{Coupon2018,Bosch2018,Aihara2022}. 

Consequently, we were left with 366 and 8317 quasars at $z=$ 2.34--3.00 in the DUD and Wide layers of 35 and 890 deg$^2$, respectively (table~\ref{tab1}).
We here removed 290 duplications in the Wide layer sample with those in the DUD layer.
We cut out $gri$ coadd images ($200\times200$ pixel$^2$ with a pixel scale of 0.168 arcsec) for selected quasars from the HSC-SSP database and conducted the spatial $2\times2$ binning to increase the signal-to-noise ratio, i.e., each image adopted in this study has $100\times100$ pixel$^2$ with a pixel scale of 0.336 arcsec.
Their identification numbers and sky coordinates from \citet{Aihara2022}, and pipeline and Ly$\alpha$ redshifts ({\tt Z\_PIPE} and {\tt Z\_LYA} from \citealt{Lyke2020}) are summarised in table~\ref{tab2}.


\section{Broad-band selection}
\label{s3}

This section discusses the details of our broad-band selection to extract extended Ly$\alpha$ emission associated with the quasar targets.
The method is conceptually the same as a well-known narrow-band technique. 
However, we employ only the broad-band ($gri$) filters: (1) we construct a continuum $g$-band image ($RI_z$) using the $r,i$-bands, and (2) subtract the extrapolated $RI_z$ image from the observed $g$-band image ($G$), which has both continuum + Ly$\alpha$ emission at $z=$ 2.34--3.00.
Here, the subscript $z$ means redshift not $z$-band.
\citet{Prescott2012} also tested a similar approach using $B,R$ bands over the 9.4 deg$^2$ Bo\"otes field.
They then confirmed that the methodology worked well by follow-up spectroscopic observations \citep{Prescott2013}.
We should note that this work fully ignores potential contributions from rest-UV helium and metal lines (e.g., \citealt{Cai2017,Guo2020}) to the broad-band images, which could cause over- or under-estimates of Ly$\alpha$ emission at some level depending on the source redshift.

\begin{figure}
\centering
\includegraphics[width=7.5cm]{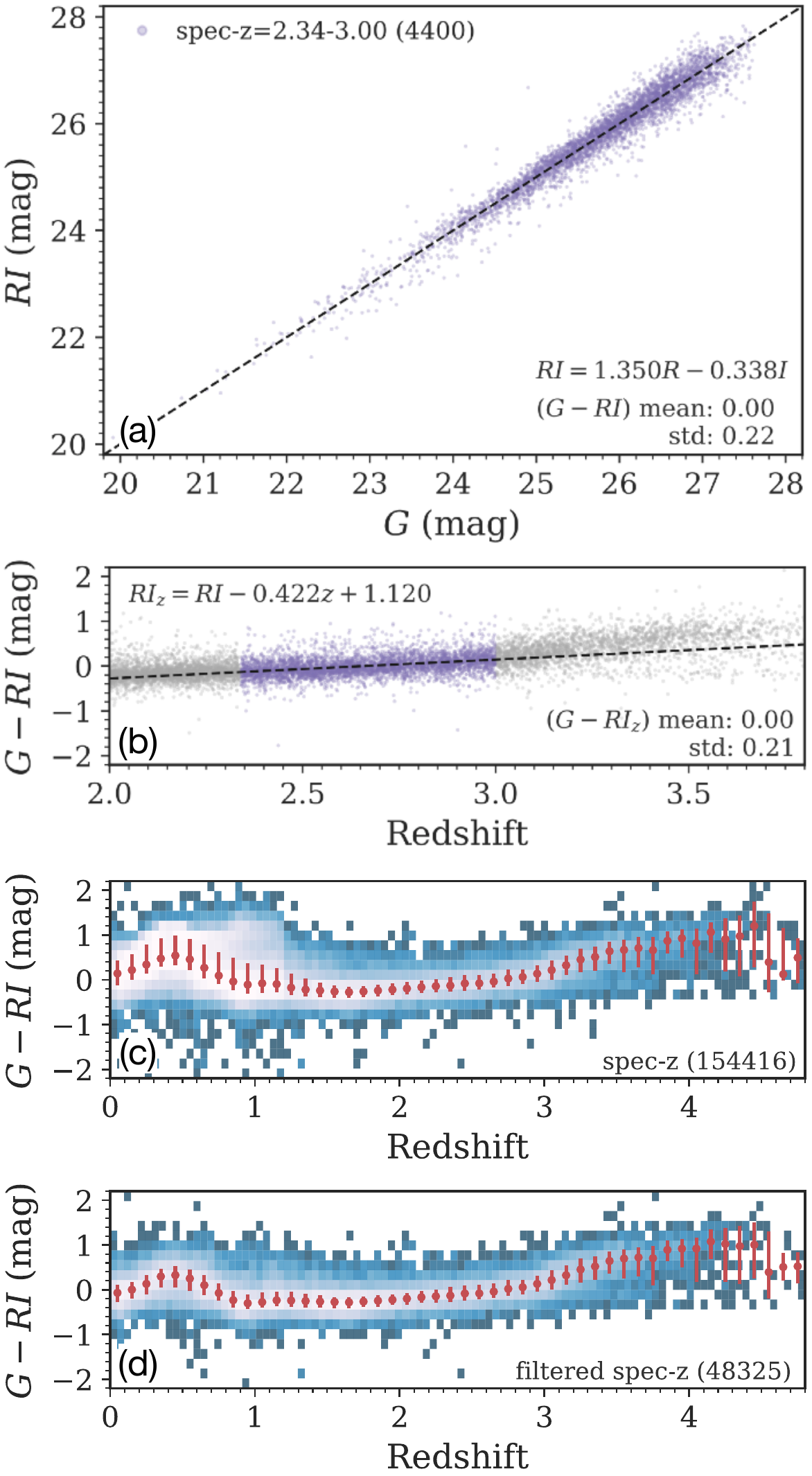}
\caption{
(a) Top panel shows comparisons between $g$-band magnitudes ($G$) and extrapolated $g$-band ($RI$) magnitudes from $r,i$-bands ($RI=1.350R-0.338I$) for 4400 spec-$z$ sources at $z=$ 2.34--3.00 in the HSC-SSP DUD layer. 
The SDSS/eBOSS quasars are not included here. 
The dashed line depicts the identity line. 
(b) $G-RI$ versus spectroscopic redshifts.
The purple dots are 4400 spec-$z$ sources at $z=$ 2.34--3.00 and 5780 spec-$z$ samples at $z<2.34$ and $3.00<z$.
The black dashed line is the best-fit line ($G-RI=0.422\ z-1.120$) for spec-$z$ sources at $z=$ 2.34--3.00.
(c) Same as the middle panel but for 154,416 spec-$z$ sources at $z=$ 0--5.
The red dots and vertical lines depict the median values and 68th percentiles at each redshift bin.
(d) Bottom panel shows the same as in (c), but with the object masking through the SVMs classification (see text).
}
\label{fig2}
\end{figure}

\begin{figure}
\centering
\includegraphics[width=8.5cm]{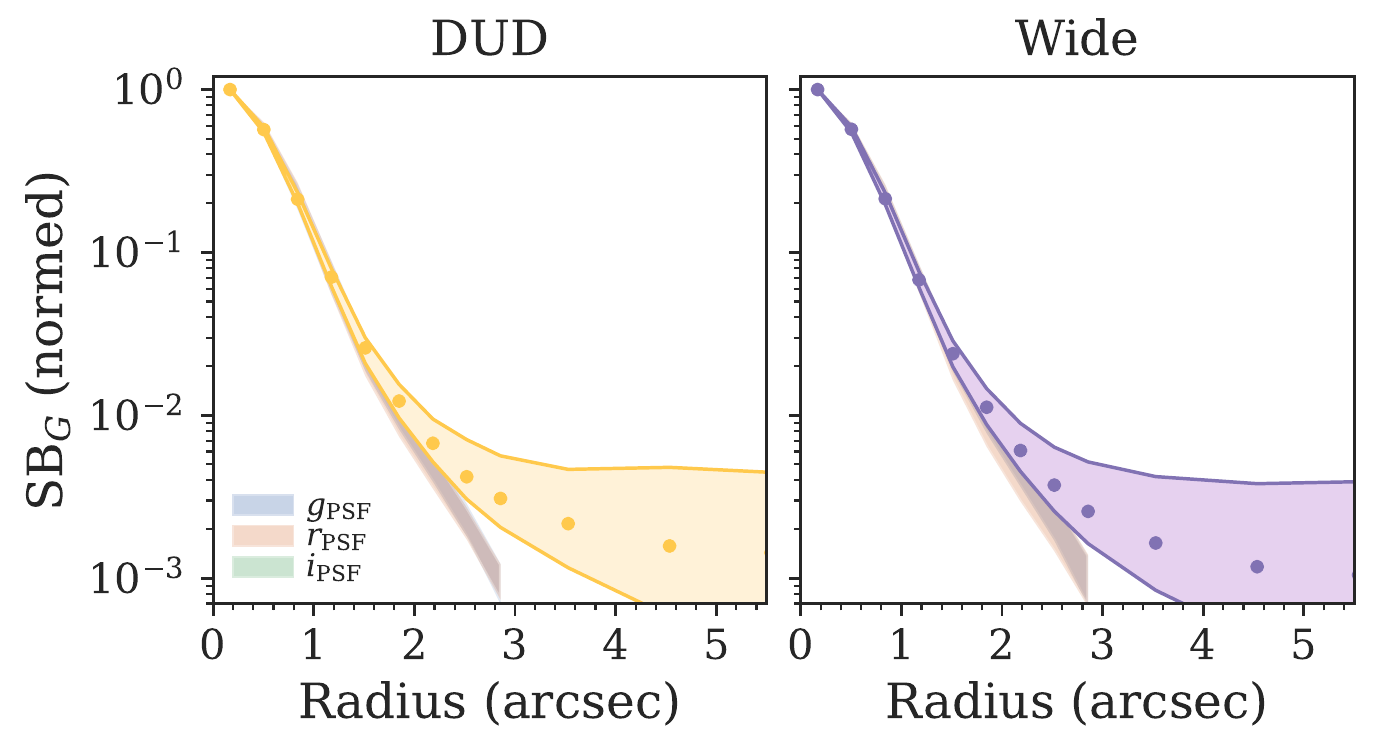}
\caption{
Radial profiles of normalised $g$-band surface flux densities of quasars at $z=$ 2.34--3.00 in the DUD layer (left) and the Wide layer (right).
The circles and colour-filled regions depict median values and 68th percentiles, respectively.
Seeing-matched PSFs (68th percentiles over each survey layer) in the $gri$-bands are also shown by blue, red, and green lines, which are mostly overlapped in each panel.
}
\label{fig3}
\end{figure}

\begin{figure}
\centering
\includegraphics[width=8.5cm]{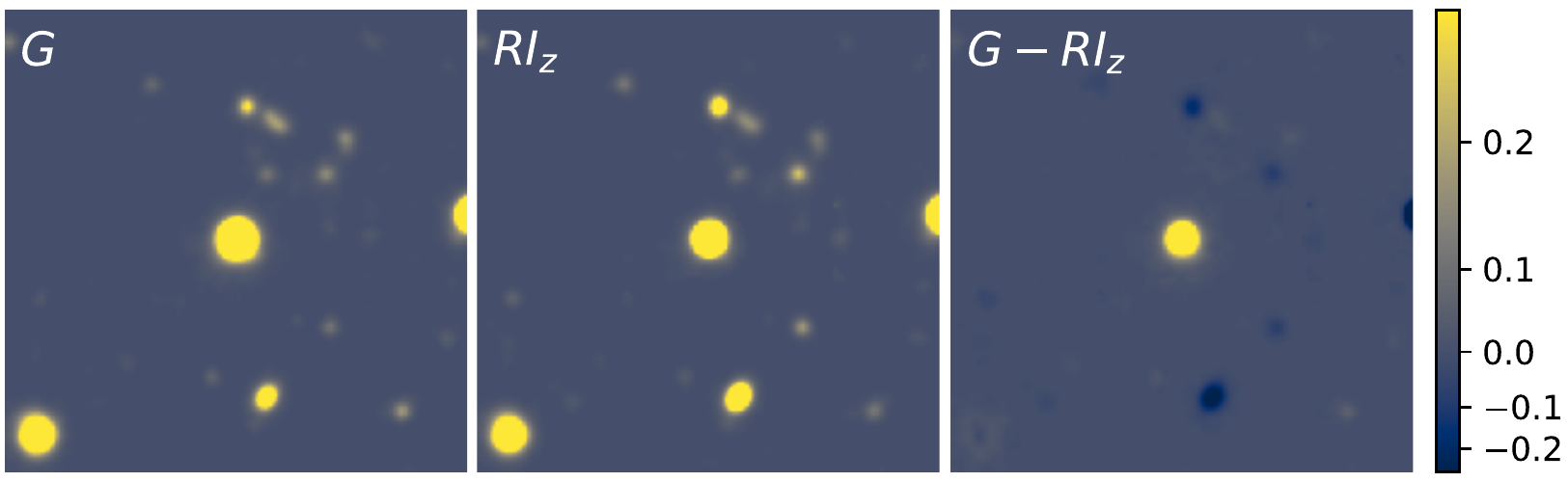}
\caption{
Examples of $G$, $RI_z$, and $G-RI_z$ images (a quasar at $z=2.85$). 
The image sizes are $34\times34$ arcsec$^2$, where the seeing FWHM are matched to 1 arcsec.
They are stretched by hyperbolic sine with the same min-max values for the sake of visibility (the colour-bar shows the pixel counts in the magnitude zero point of 27 mag).
}
\label{fig4}
\end{figure}

For extracting the extended Ly$\alpha$ emission, it is most important to subtract the rest-frame ultra-violet continua of host galaxies and satellites.
At the same time, we do not consider quasars themselves because they can be regarded as point sources.
At first, an empirical extrapolation of the $RI$-band images from the $r,i$-bands ($R,I$) is performed based on 4400 spec-$z$ sources at $z=$ 2.34--3.00 from 3D-HST \citep{Brammer2012,Momcheva2016}, DEIMOS 10k sample \citep{Hasinger2018}, PRIMUS \citep{Coil2011,Cool2013}, and VVDS \citep{LeFevre2013},
\begin{equation}
    RI = 1.350\ R - 0.338\ I.
\label{eq1}
\end{equation}
The best-fit relation (figure~\ref{fig2}) is obtained using a non-linear optimisation and curve-fitting tool for Python, {\tt lmfit} \citep{Newville2014}. 
We do not incorporate the SDSS/eBOSS quasars into the spec-$z$ sample here.
In practice, Ly$\alpha$ emission of these spec-$z$ sources should also affect $g$-band photometry.
However, we assume that their average Ly$\alpha$ emission does not significantly affect the fitting.
We indeed confirm that the best-fit relation well traces spec-$z$ sources at $z<2.3$, where there is no Ly$\alpha$ contribution to the $g$-band (figure~\ref{fig2}).
Besides, additional colour correction has been applied to calibrate a small redshift ($z$) dependence as seen in figure~\ref{fig2}b,
\begin{equation}
    RI_z = RI - 0.422\ z + 1.120,
\label{eq2}
\end{equation}
where $RI_z$ is the obtained continuum image adopted in this work.
We matched seeing sizes of all the $gri$ imaging data to FWHM $=1$ arcsec of the original sizes obtained from the adaptive moments of the PSF model ({\tt sdssshape\_psf\_shape}; see \citealt{Aihara2022,Bernstein2002}) by Gaussian smoothing.
Seeing-matched radial profiles of surface flux densities of quasars in the $g$-band and PSF in the $gri$-bands are presented in figure~\ref{fig3}.
Galactic extinctions were also corrected for the individual $gri$ images.
We show sample $G$, $RI_z$, and $G-RI_z$ images of a target quasar at $z=2.35$ in figure~\ref{fig4}, demonstrating that we can erase continuum sources in the residual $G-RI_z$ map.
The residual map is critical for this study to search for extended Ly$\alpha$ emissions around the quasar sample as described in section~\ref{s4}.
However, we should note that some clear over- and under-subtractions can be observed due to different colour-terms of foreground and background sources, as discussed in detail below.

Figure~\ref{fig2} summarises the empirical extrapolations of $g$-band without Ly$\alpha$ contributions.
The derived relation (eq.~\ref{eq1}) well extrapolates $g$-band photometry of spec-$z$ sources from those in the $r,i$-bands with a standard deviation of 0.22 mag, corresponding to $\sim20$\% errors in the residual Ly$\alpha$ fluxes.
The further redshift correction (eq.~\ref{eq2}) marginally improves the extrapolation.
We observed an increase of scatter of $G-RI$ at $z>3$, likely caused by significant contributions of Ly$\alpha$ and Ly$\beta$ forests.
Given such an increasing scatter at $z>3$, we did not adopt quasars at $z=$ 3.0--3.5, although their Ly$\alpha$ lines fell into the $g$-band data.
Figure~\ref{fig4}c shows the $G-RI$ distributions in a wider redshift range ($z=$ 0--5), where we employed spec-$z$ sources from GAMA DR3 \citep{Baldry2018} and SDSS DR15 \citep{Aguado2019}.\footnote{Although the data release paper \citep{Aihara2022} refers to different versions in GAMA (DR2) and SDSS (DR16), release versions cited here are the right versions.}
Increasing colour scatter can be found at $z<1.5$ and $z>3$, requiring us to be aware of irrelevant neighbours, especially outliers in the foreground.
Otherwise, we may overestimate an area of extended Ly$\alpha$ emission.

\begin{figure}
\centering
\includegraphics[width=7.5cm]{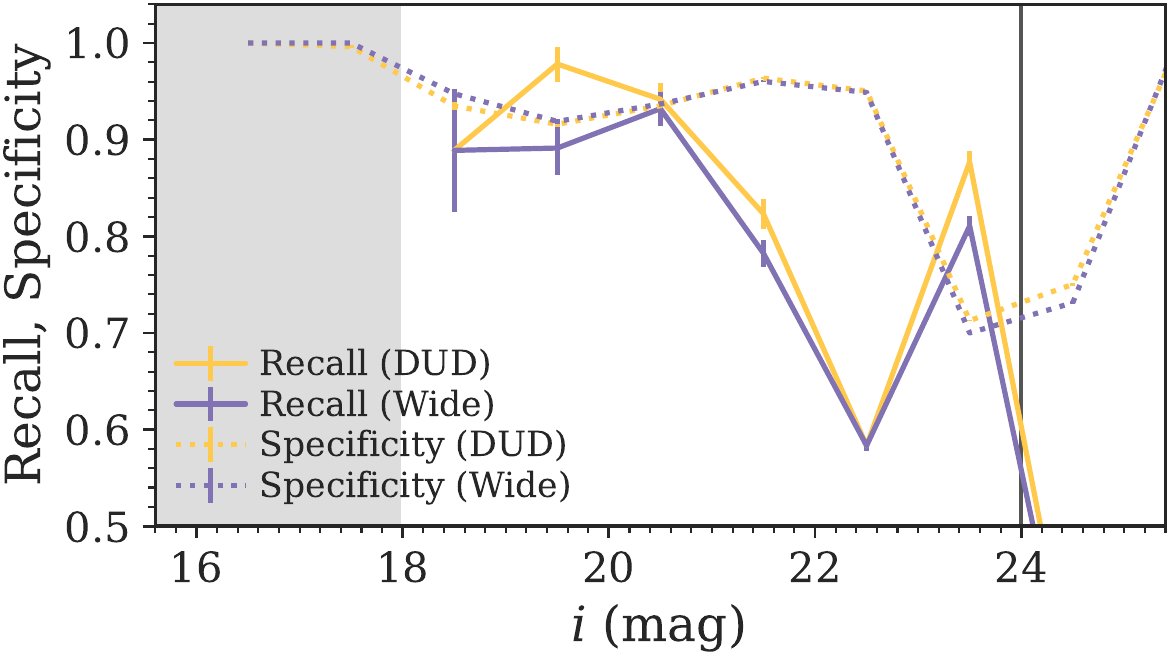}
\caption{
Mean values and $1\sigma$ Poisson errors of recall and specificity rates in the SVMs classification for spec-$z$ sources at $z=$ 2.34--3.00 ($N=3559$, the solid line) and at $z<2.34$, $3.00<z$ ($N=135569$, the dotted line), respectively, as a function of $i$-band magnitude ($\Delta=0.1$ mag). 
We adopted identical spec-$z$ sources in the DUD and Wide layers, which are represented by the yellow and purple lines, respectively. 
There is no $z=$ 2.34--3.00 sample at $i<18$ (the grey area).
We did not apply the object masking for faint sources ($i<24$, the vertical line), where we confirmed low recall rates due to weak detection and a lack of the training samples.
}
\label{fig5}
\end{figure}

To minimise such contaminants from the foreground and background neighbours, we masked outlier candidates at $z<2.34$, $3.00<z$ based on the $grizy$ magnitudes ({\tt cmodel}; \citealt{Abazajian2004,Bosch2018}) with support vector machines (SVMs; \citealt{Boser1992,Cortes1995}).
We employed {\tt Scikit-learn} (version 0.24.2; \citealt{scikit-learn}) to implement the SVMs classification with Radial Basis Function (RBF) kernel.
First, we established two spec-$z$ samples at the target redshifts of $z=$ 2.34--3.00 ($N=746$) and at foreground or background redshifts $z<2.34$, $3.00<z$ ($N=105606$) covered by both the DUD and Wide layers.
We here applied the $i$-band magnitude cut ($i<24$ mag) since the number of spec-$z$ references significantly drops at $i>24$ mag.
Using $grizy$ photometry of $746+105606$ training samples, we obtained the best decision boundaries in five-dimensional space to classify into these two populations through SVMs with weight by the sample sizes.
The derived total accuracy scores were 0.893 in the DUD layer and 0.889 in the Wide layer.
We then evaluated recall and specificity rates (so-called completeness) in different $i$-band magnitudes as seen in figure~\ref{fig5}, which are respectively defined by TP/(TP+FN) and TN/(TN+FP).
TP (or TN) and FN (or FP) are true positive (or negative) and false negative (or positive) values, where the positive and negative classes indicate the sources at $z=$ 2.34--3.00 and $z<2.34$, $3.00<z$ predicted by the SVMs classifier, respectively.

Consequently, we found that the SVMs classifier can select $z=$ 2.34--3.00 and $z<2.34$, $3.00<z$ sources with on average $\sim80$\% and $\sim90$\% completeness down to $i=24$ mag (figure~\ref{fig5}).
Based on the classification result, we masked foreground or background neighbours with $i=$ 18--24 mag around the target quasars to minimise the colour-term contaminants.
We observed that this masking process can significantly improve the colour-term effect at $z<1.5$ (figure~\ref{fig2}d), though it remains imperfect due to the limited photometric information.
The mask areas were determined by three times PSF-convolved major- and minor-axis based on the second moments of the object intensity, termed adaptive moments in the $i$-band ({\tt i\_sdssshape\_shape}; see \citealt{Bernstein2002}).
We also masked bright objects with $i<18$ mag and point sources ({\tt i\_psfflux\_mag$-$i\_cmodel\_mag$<0.2$}; see \citealt{Strauss2002,Baldry2010}).
Additionally, quasar itself was masked within 2.5 arcsec diameter ($2.5\times$ seeing FWHM or $r\lesssim10$ pkpc) for relatively fair comparisons of Ly$\alpha$ properties between optically bright and faint quasars (section~\ref{s5}).
Examples of masked $G-RI_z$ images are shown in figure~\ref{fig6} and \ref{fig8} in section~\ref{s4}.


\section{Quasars with Lyman-Alpha nebulae}
\label{s4}

Up to this point, we have explained how we selected quasars and established Ly$\alpha$ images around the targets.
This section reviews the broad-band selection for 8683 quasars at $z=$ 2.34--3.00, and the search for extended Ly$\alpha$ emission around the targets to select quasars with large gaseous nebulae (hereafter called \emph{quasar-nebulae}).

\begin{figure*}
\centering
\includegraphics[width=15cm]{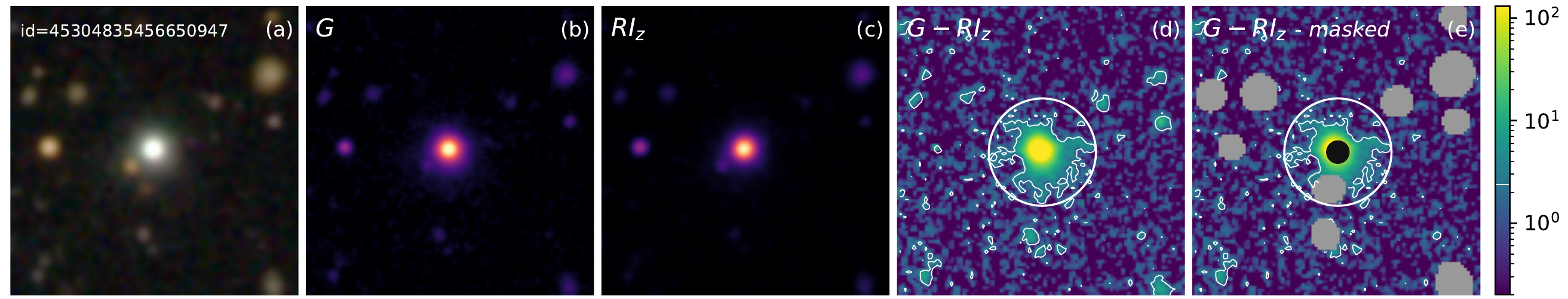}
\caption{
From the left to the right, (a) $gri$ colour image, (b) $g$-band coadd image, (c) extrapolated $g$-band ($RI_z$) image, and (d,e) residual $G-RI_z$ image without and with masks, respectively, for a known quasar-nebula, SDSS~J1025+0452 at $z=3.2$ \citep{ArrigoniBattaia2019}, which is covered by the HSC-SSP PDR3 Wide layer. 
The images have a length and width of 34 arcsec ($\sim260$ pkpc). 
The colourbar indicates the tentatively-measured Ly$\alpha$ surface densities ($10^{-17}$ erg~s$^{-1}$cm$^{-2}$arcsec$^{-2}$) by eq.~\ref{eq3}.
The white contour in the residual map depicts $2\sigma$ excess from the background deviation ({\tt bg\_std} but with the filter correction). 
The white open circles indicate the maximum radius for calculating {\tt Area} and {\tt Area\_eff} ($=55.32$ and 46.17 arcsec$^2$, respectively).
Although the source redshift is a bit out of the target redshift range $z=$ 2.34--3.00, the residual $G-RI_z$ image reproduced the extended Ly$\alpha$ nebula seen in the deep VLT/MUSE datacube by \citet[\#10 of figure~1 and figure~5]{ArrigoniBattaia2019}. 
The grey regions and the black centre circle in the right panel (f) are the object masks for foreground or background contaminants (see text) and quasar itself (2.5 arcsec diameter or $r\lesssim10$ pkpc), respectively.
}
\label{fig6}
\end{figure*}

We first tested our procedure to a known quasar-nebula source, SDSS~J1025+0452 at $z=3.2$, covered by the HSC-SSP PDR3 Wide layer (figure~\ref{fig6}).
The QSO MUSEUM survey reported this quasar-nebula \citep{ArrigoniBattaia2019}, extending across 289 arcsec$^2$ in the deep VLT/MUSE datacube (surface brightness limit of $4.2\times10^{-18}$ erg~s$^{-1}$cm$^{-2}$arcsec$^{-2}$ in 1 arcsec$^2$ for the 30 \AA\ NB image according to the literature).
We stress that this source redshift is slightly off from our target redshift slice, $z=$ 2.34--3.00.
Therefore, it may have higher extinction in the $g$-band than our assumption due to IGM absorption (figure~\ref{fig2}).
Nevertheless, the resultant residual $G-RI_z$ image replicated extended Ly$\alpha$ emission traced by \citet[figure~1 and 5]{ArrigoniBattaia2019}, whereas our sensitivity-limited broad-band selection could detect Ly$\alpha$ signals only up to an area of 55.32 arcsec$^2$ (or 46.17 arcsec$^2$ with masks).

\begin{figure}
\centering
\includegraphics[width=8.5cm]{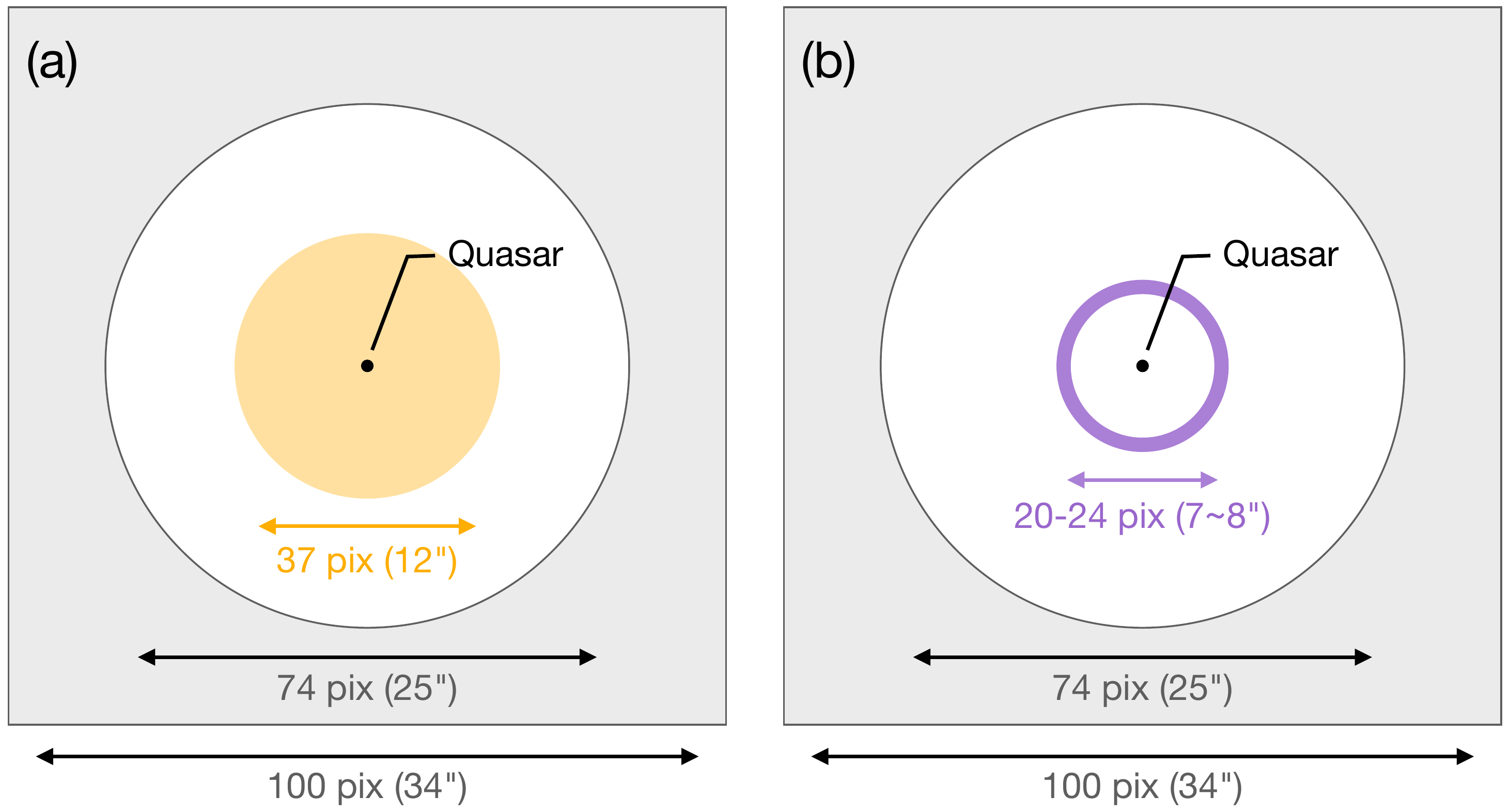}
\caption{
Schematic illustration of the selection procedure. 
(a) We calculate a pixel area ({\tt Area}) with $>2\sigma$ excess of residual flux within a radius of 6 arcsec from the centre (the orange circle). 
The background deviation is estimated at $r>25$ arcsec ($\gtrsim100$ pkpc; the grey region) for each $G-RI_z$ cutout image. 
This work selects quasars with {\tt Area\_eff} $>40$ arcsec$^2$ ($>442$ pixel$^2$) as quasar-nebulae. 
(b) We also derive the mean surface brightness ({\tt SB\_ann}; eq.~\ref{eq3}) in an annulus of radius 3.5--4.0 arcsec (corresponding to $\sim30$ pkpc; the purple area), which may be useful for users to make a strategy of follow-up observation. 
}
\label{fig7}
\end{figure}

We now introduce our selection of the quasar-nebulae.
Throughout this paper, quasars with effective Ly$\alpha$ areas ({\tt Area\_eff}) greater than 40 arcsec$^2$ are deemed to be quasar-nebulae, where {\tt Area\_eff} is defined as the pixel area with $2\sigma$ excess from the background deviation ({\tt bg\_std}), excluding the masked area (figure~\ref{fig6} and \ref{fig7}).
Directly obtained Ly$\alpha$ areas without masking irrelevant sources are defined as {\tt Area}.
Both {\tt Area} and {\tt Area\_eff} were calculated within a 37 pixel ($\sim12$ arcsec or $\sim100$ pkpc) aperture diameter, where the aperture limit is set to reduce the effects from the foreground and background neighbours.
The diameter size is also sufficiently smaller than the mesh size (43 arcsec) for the local sky subtraction in the image processing by HSC-SSP \citep{Aihara2022}.
The background deviation {\tt bg\_std} was estimated based on pixels at $r>37$ pixels (figure~\ref{fig7}).
The obtained residual Ly$\alpha$ maps reached typical $2\sigma$ depths per $2\times2$ binned pixel down to $\sim0.9\times10^{-17}$ erg~s$^{-1}$cm$^{-2}$arcsec$^{-2}$ in the DUD layer and $\sim1.6\times10^{-17}$ erg~s$^{-1}$cm$^{-2}$arcsec$^{-2}$ in the Wide layer, respectively.
We also tentatively estimated the Ly$\alpha$ surface brightness at $r=$ 10--12 arcsec ({\tt SB\_{ann}}; figure~\ref{fig7}), defined as,
\begin{equation}
    \mathtt{SB\_ann} = \Delta_g w_G(z)~\mathrm{SB}_{G-RI_z}.
\label{eq3}
\end{equation}
It may be convenient for users to infer the expected Ly$\alpha$ surface brightness in the outskirts ($r\sim30$ pkpc).
The sign $\Delta_g$ is FWHM of the $g$-band filter ($=1375$ \AA), and $w_G(z)$ is an weight parameter for correcting the $g$-band transmittance (figure~\ref{fig1}), which yields, e.g., 1.79 at $z_\mathrm{Ly\alpha}=2.34$ and 0.98 at $z_\mathrm{Ly\alpha}=3.00$.
SB$_{G-RI_z}$ is mean surface brightness of the residual Ly$\alpha$ image ($G-RI_z$) at $r=$ 10--12 arcsec.
One should note that {\tt SB\_{ann}} values vary depending on many factors such as imaging depths, contamination by projected neighbours in the foreground and background, and colour term effects.
Therefore, they should be used \emph{only as a guide}, e.g., for making a strategy of a follow-up observation.

\begin{figure*}
\centering
\includegraphics[width=15.5cm]{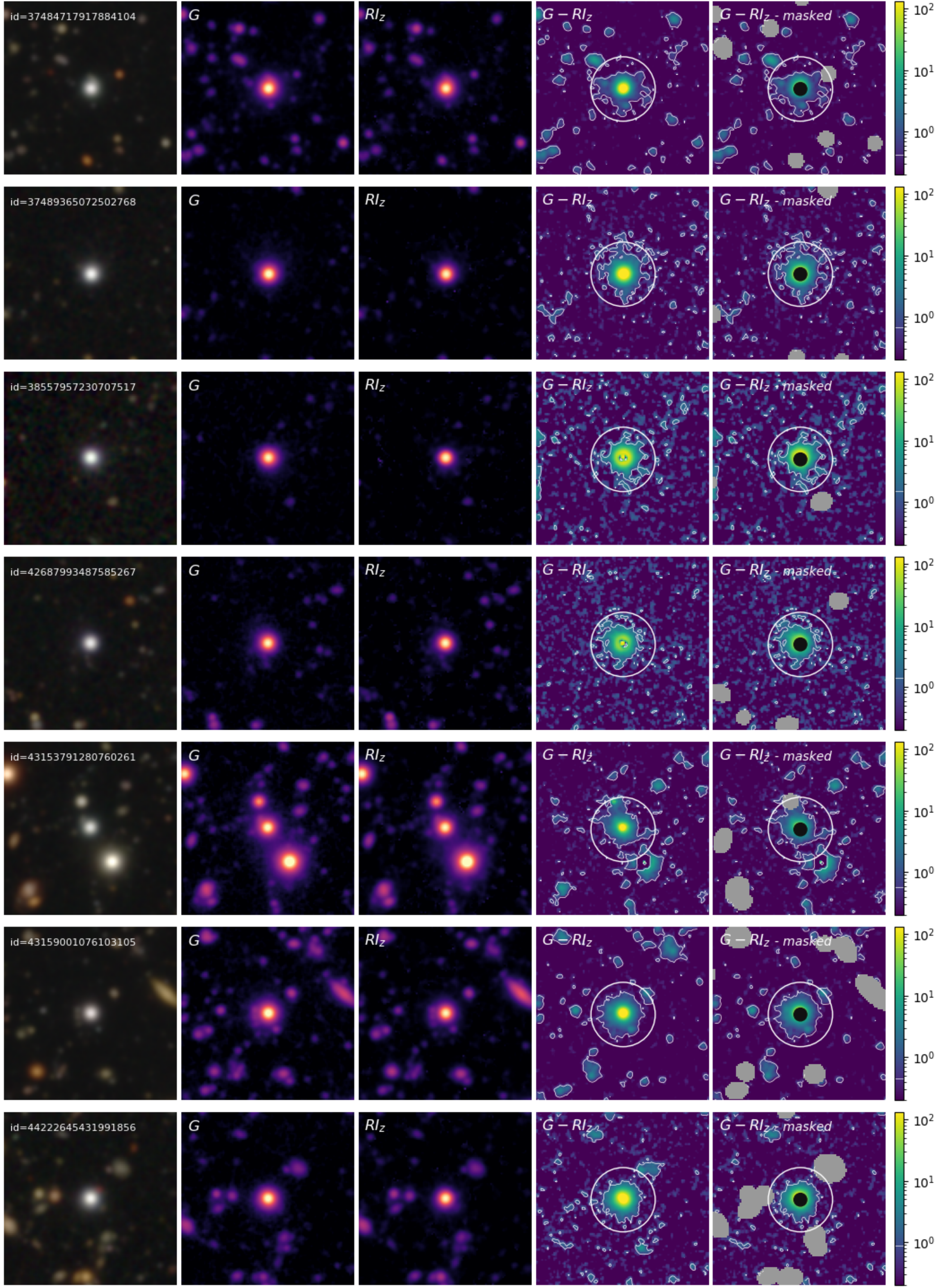}
\caption{
Same as figure~\ref{fig6}, but seven quasar-nebula candidates in the HSC-SSP PDR3 DUD layer, all of which have {\tt Area\_eff} $>40$ arcsec$^2$. 
The colourbars indicate tentatively-measured Ly$\alpha$ surface brightness in $10^{-17}$ erg~s$^{-1}$cm$^{-2}$arcsec$^{-2}$ (eq.~\ref{eq3}).
Those in the Wide layer are available in figure~\ref{fig1a}. 
}
\label{fig8}
\end{figure*}

We applied these measurements to all quasar targets and then searched for quasar-nebulae with extended Ly$\alpha$ emission ({\tt Area\_eff} $>40$ arcsec$^2$).
The source catalogue available as online material (table~\ref{tab2}) summarises obtained {\tt Area}, {\tt Area\_eff}, {\tt SB\_{ann}}, and {\tt bg\_std} for 366 and 8317 quasars at $z=$ 2.34--3.00 in the HSC-SSP PDR3 DUD and Wide layers (table~\ref{tab1}).
With the selection threshold of {\tt Area\_eff} $>40$ arcsec$^2$, we respectively select 7 and 32 sources as quasar-nebulae.
Processed images like figure~\ref{fig6} for all quasar-nebulae in the DUD and Wide layers are represented in figure~\ref{fig8} and \ref{fig1a}, respectively, showing clear extended Ly$\alpha$ features in both unmasked and masked $G-RI_z$ images.
These processed images help catalogue users to visually check credibility of residual Ly$\alpha$ distributions from the broad-band selection.
Refer to the data availability section for more details on acquiring the cutout data.
If one needs to access the risk catalogue for $z=$ 3.0--3.5 quasars unused in this study, please contact the author.

\begin{table*}
\caption{
Quasar information and properties of Ly$\alpha$ emission. 
The full source catalogue is available as online material in CSV format.
}
\label{tab2}
\begin{tabular}{lllllllrrrr} 
	\hline
	{\tt object\_id} & layer & depth$^a$ & ra & dec & {\tt Z\_PIPE}$^b$ & {\tt Z\_LYA}$^b$ & {\tt Area}$^c$ & {\tt Area\_eff}$^c$ & {\tt SB\_ann$^d$} & {\tt bg\_std}$^d$\\
	\hline
	37484563299063297 & dud & 28.16 & 34.93038 & -5.43555 & 2.78811 & 2.85555 & 19.19 & 14.22 & 1.20 & 2.39\\
	36429328489138776 & dud & 26.67 & 36.21685 & -6.01799 & 2.91488 & 2.91186 & 22.47 & 13.32 & 6.88 & 8.73\\
	37484567594046770 & dud & 28.31 & 34.95245 & -5.12826 & 2.54855 & 2.54530 &  4.40 &  3.84 & 0.91 & 2.01\\
	37484717917884104 & dud & 28.36 & 34.83848 & -4.72088 & 2.77543 & 2.80256 & 59.04 & 54.08 & 4.95 & 1.78\\
	37484705032993293 & dud & 28.32 & 34.85126 & -5.19717 & 2.53069 & 2.52831 & 15.13 & 10.16 & 2.61 & 2.74\\
    \multicolumn{11}{c}{... (366 rows in total)}\\
	36411452835251556 & wide & 26.53 & 30.55460 & -6.38877 & 2.53873 & 2.53747 &  8.47 &  4.40 & 4.31 & 7.52\\
    \multicolumn{11}{c}{... (8317 rows in total)}\\
	\hline
\multicolumn{11}{l}{$^a$ $5\sigma$ PSF limiting magnitude in the $g$-band from \citet{Aihara2022}}\\
\multicolumn{11}{l}{$^b$ Excerpts from \citet{Lyke2020}}\\
\multicolumn{11}{l}{$^c$ arcsec$^2$}\\
\multicolumn{11}{l}{$^d$ $10^{-18}$ erg~s$^{-1}$cm$^{-2}$arcsec$^{-2}$}\\
\end{tabular}
\end{table*}

\section{Discussion: Wheat versus Chaff in Lyman-Alpha emission around quasars}
\label{s5}

The broad-band colour selection enabled us to extract Ly$\alpha$ emission around quasars at $z=$ 2.34--3.00.
We successfully obtained 7 and 32 quasar-nebulae with effective Ly$\alpha$ areas greater than 40 arcsec$^2$.
This section investigated what factors distinguish the spatial extent of Ly$\alpha$ emission between \emph{wheat and chaff}, i.e., quasars with and without large Ly$\alpha$ nebulae, by examining {\tt Area\_eff} distributions of control samples with respect to different parameters.
Specifically, motivated by the previous discussions about the limited quasar samples (see, e.g., \citealt{ArrigoniBattaia2019,Cai2019}), we investigated connections of the extent of Ly$\alpha$ emission to absolute magnitude and radio-loudness.

\begin{figure}
\centering
\includegraphics[width=7.5cm]{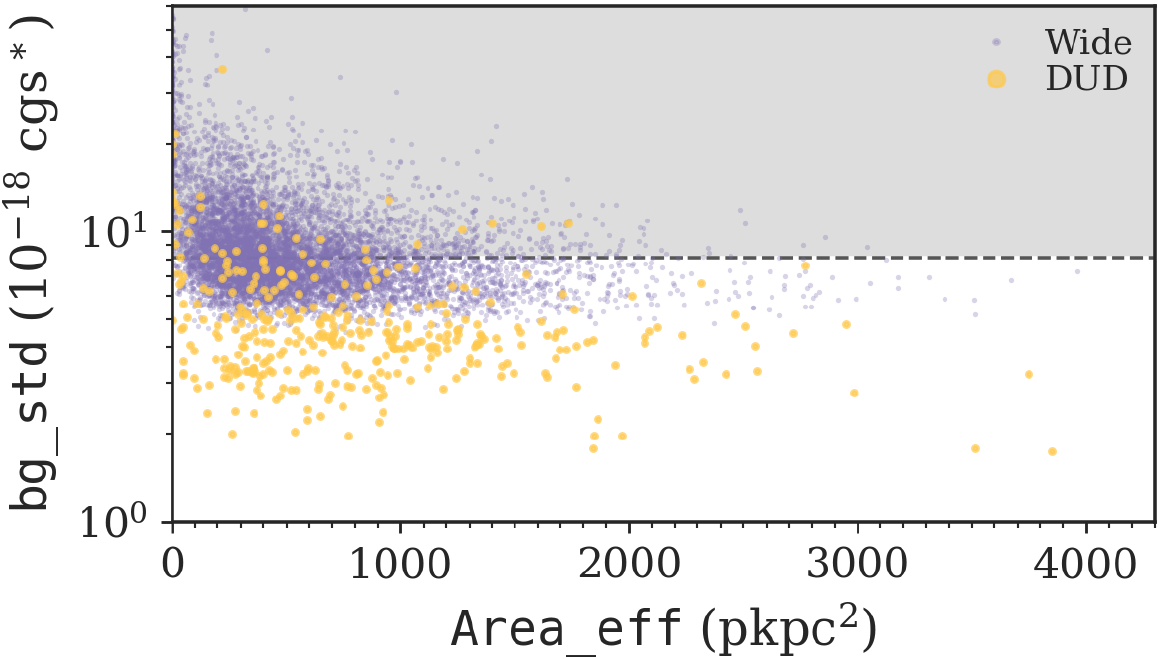}
\caption{
Background deviation {\tt bg\_std} ($\ast$ erg~s$^{-1}$cm$^{-2}$arcsec$^{-2}$) versus effective Ly$\alpha$ area {\tt Area\_eff} for our samples in the DUD (yellows) and Wide (purples) layers.
For fair comparisons, we excluded quasars with relatively shallower data ({\tt bg\_std} $>8.127$) throughout the discussion section.
The threshold was determined by 90\% completeness limit of the quasar-nebula sample in the Wide layer.
}
\label{fig9}
\end{figure}

We first removed quasars with relatively poor imaging qualities inferred from figure~\ref{fig9}, which showed the background deviation in pixel ({\tt bg\_std}) as a function of the effective Ly$\alpha$ area {\tt Area\_eff}.
We here changed the unit of {\tt Area\_eff} from arcsec$^2$ to pkpc$^2$ to correct the small redshift difference at $z=$ 2.34--3.00.
We set the selection threshold {\tt bg\_std} $<8.127\times10^{-18}$ erg~s$^{-1}$cm$^{-2}$arcsec$^{-2}$, corresponding to the 90\% completeness of quasar-nebulae in the Wide layer.
Given the selection criterion, we established the clean samples of 328 and 4212 quasars in the DUD and Wide layers, respectively.
Additionally, to check the radio detection, we cross-matched our quasar sample with the FIRST survey catalogue (version 14Dec17; \citealt{Becker1995,White1997}) within a radius of 1 arcsec.
The FIRST survey covers the entire HSC-SSP field, and the source catalogue contains radio sources with the detection limit of 1 mJy at $\sim1.4$ GHz.
This paper defined 4 and 112 spatially-matched quasars as radio-loud quasars in the DUD and Wide layers, respectively, and the others as radio-quiet quasars.
We note that radio-quiet quasars defined here could be still radio-loud quasars.

\begin{figure}
\centering
\includegraphics[width=8cm]{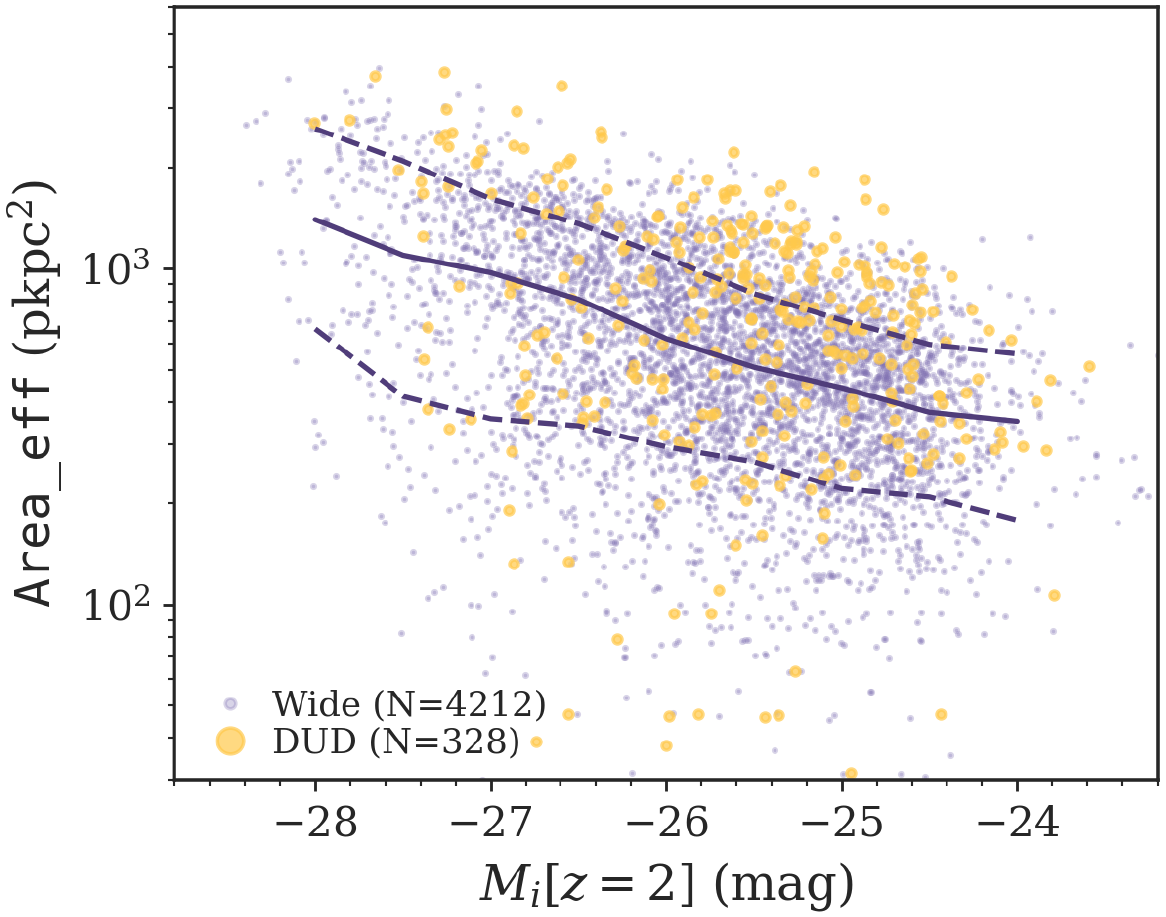}
\caption{
Effective Ly$\alpha$ areas ({\tt Area\_eff}) versus absolute $i$-band magnitudes normalised at $z=2$ ($M_i[z=2]$ from \citealt{Lyke2020}) for clean samples in the DUD (yellows) and Wide (purples) layers.
The purple solid and dashed lines depict median values and 68th percentiles given $M_i[z=2]$ bins ($\Delta=0.25$ mag) for the Wider layer sample.
}
\label{fig10}
\end{figure}

Figure~\ref{fig10} compares {\tt Area\_eff} with absolute $i$-band magnitudes normalised at $z=2$ ({\tt M\_I} from \citealt{Lyke2020} with the K corrections in \citealt{Richards2006}, hereafter $M_i[z=2]$) for the clean samples in the DUD and Wide layers.
According to the Spearman's rank correlation test, we detected moderate correlation between effective Ly$\alpha$ area ({\tt Area\_eff}) and $M_i[z=2]$ by correlation coefficients $r_s=-0.28$ in the DUD layer ($N=328$) and $r_s=-0.44$ in the Wider layer sample ($N=4212$) with sufficiently small $p$-values ($p\ll0.001$).
The moderate correlations suggest that more luminous quasars tend to involve more extended Ly$\alpha$ emission with high Ly$\alpha$ surface brightness $\gtrsim1\times10^{-17}$ erg~s$^{-1}$cm$^{-2}$arcsec$^{-2}$.

Such observed tendencies are broadly consistent with results from previous surveys to $z>2$ quasars.
\citet{ArrigoniBattaia2019,Mackenzie2021} have reported that more luminous quasars (i.e., lower $M_i[z=2]$) have higher peak Ly$\alpha$ luminosity.
The currently known ELANe are mostly associated with the brightest quasars at $M_i\lesssim-27$ mag \citep{Cantalupo2014,Hennawi2015,Cai2017,Cai2018,ArrigoniBattaia2018}.
A deep narrow-band Ly$\alpha$ imaging survey also detected Ly$\alpha$-illuminated cosmic web structures around a hyper-luminous quasar at $z=2.84$ \citep{Kikuta2019}.
Combined with these previous findings, the results indicate at least that there is a higher chance to detect very extended and luminous Ly$\alpha$ emission around the brightest quasars, i.e., more massive black holes \citep{McLure2004,Kollmeier2006}, within reasonable observation times, as seen in the higher detection rate of quasar-nebulae in more luminous samples.
In other words, our limited-sensitivity would be missing extended but faint Ly$\alpha$ structures, especially around less luminous quasars.
Furthermore, previous studies have reported a relatively little variation in Ly$\alpha$ properties from redshift $z\sim6$ to $z\sim3$ \citep{Farina2019}, but a lower averaged surface brightness of Ly$\alpha$ nebulosities at $z=2$ compared to those at $z=3$ \citep{ArrigoniBattaia2019,Cai2019}.
This implies the lower covering factors of cold gas components in the CGM at $z<3$ than those at $z>3$, based on the assumption that the Ly$\alpha$ surface brightness is proportional to the number density of hydrogen and the column density in the optically thin environment \citep{Cantalupo2014,ArrigoniBattaia2015,ArrigoniBattaia2016,Cai2019}.
However, this paper does not delve into the redshift dependence given the limited redshift range ($z=$ 2.34--3.00) and the redshift dependent colour-term effect.
We leave such detailed work to future work.

\begin{figure}
\centering
\includegraphics[width=8.5cm]{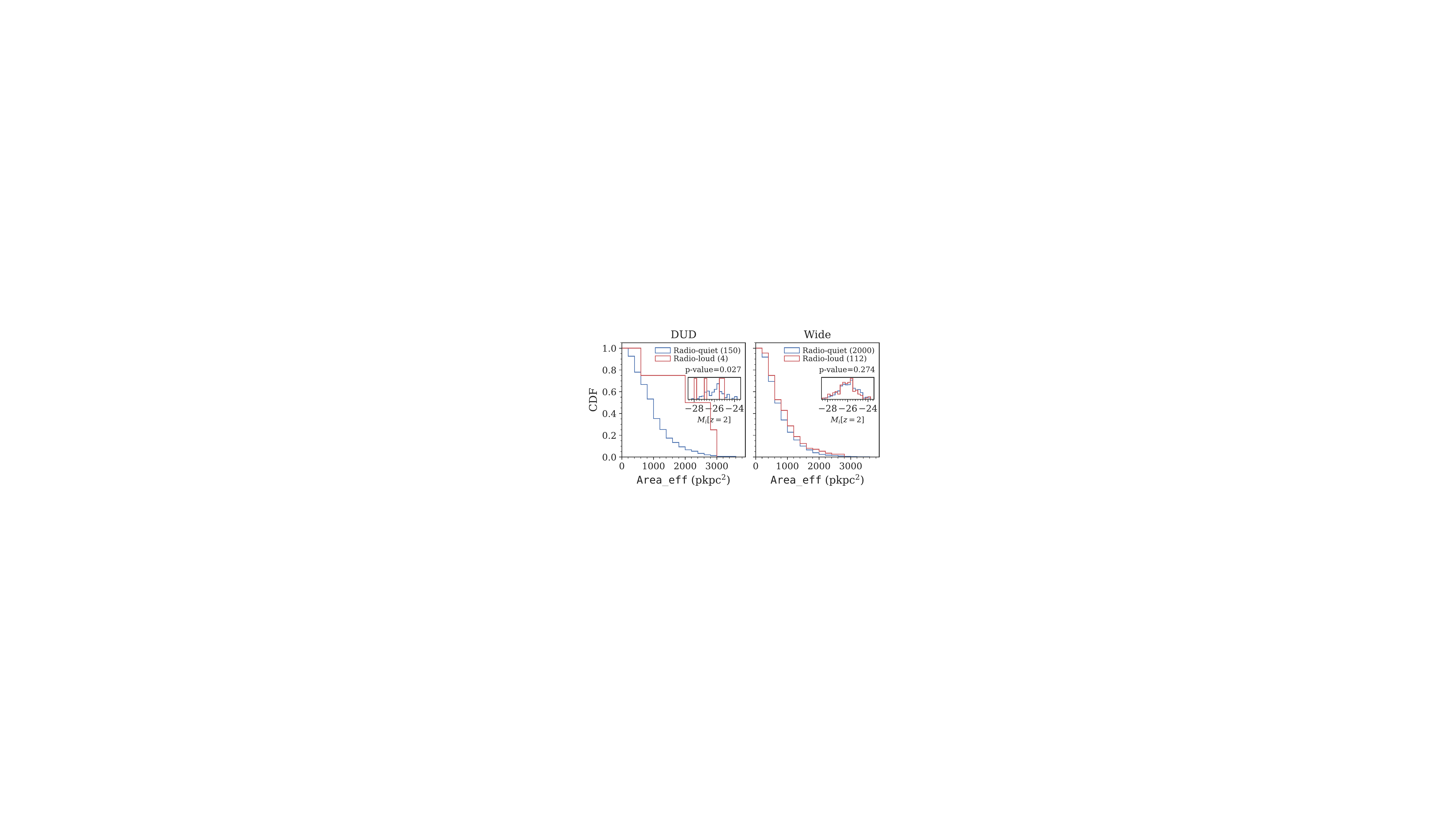}
\caption{
The obtained $p$-value from the two-sample KS tests is inserted in each panel.
The numbers in parentheses in each legend are the number of quasars used in each group.
In the middle and right panels, the comparisons of absolute $i$-band magnitude distributions ($M_i[z=2]$) for the control samples are shown.
}
\label{fig11}
\end{figure}

Next, {\tt Area\_eff} distributions for $M_i[z=2]$-matched control samples of radio-loud and radio-quiet quasars are represented in figure~\ref{fig10}.
Inset panels of figure~\ref{fig10} show that we well matched $M_i[z=2]$ values in the Wide layer, while we lacked radio-loud quasars in the DUD layer ($N=4$).
The two-sample Kolmogorov--Smirnov (KS) test indicates no significant difference in distributions of the effective Ly$\alpha$ areas between radio-loud and radio-quiet quasars, as seen in $p$-value $=0.274$ (figure~\ref{fig11}), suggesting that their Ly$\alpha$ surface profile are similar in a large sense.
The systematic IFU observation of $z\sim3$ quasars by \citet{ArrigoniBattaia2019} also reported similar mean Ly$\alpha$ profiles between radio-loud and radio-quiet quasars, where only a marginal excess in radio-loud samples at a radius of 30--50 pkpc were detected.
It is thus consistent with the statistical trend obtained in our comparison samples.
Contrarily, given the same control samples, we detected a higher fraction ($5.4\pm2.2$\%) of radio-loud quasars with relatively large {\tt Area\_eff} ($>2000$ pkpc$^2$) than that of radio-quiet quasars ($2.4\pm0.4$\%).
The higher detection rate of the more extended Ly$\alpha$ emission around radio-loud quasars could be related to viewing angles of radio jets and/or evolutionary phases of radio-loud quasars, as many case studies presented spatial and kinematic interactions between radio jets and Ly$\alpha$ nebulae (e.g., \citealt{McCarthy1990,vanOjik1996,Pentericci1997,Morais2017,Silva2018}).
We tentatively checked the size measurements of radio sources associated with radio-loud quasars from \citet{Becker1995,White1997}, which are based on the elliptical Gaussian model. 
However, we did not see any systematic relation to their Ly$\alpha$ areas so far.
Assessing individual radio-loud quasars with multi-wavelength data sets would be important to delve into additional processes contributing to their extended Ly$\alpha$ emissions.
We defer such detailed analyses of individual objects to future work.

Lastly, we discuss whether or not our quasar-nebulae include the largest class of ELANe such as the Slug nebula at $z=2.3$ \citep{Cantalupo2014} and the MAMMOTH-1 at $z=2.3$ \citep{Cai2017}.
These largest ELANe have very extended Ly$\alpha$ nebulae beyond 150 pkpc with high Ly$\alpha$ surface brightness $\gtrsim1\times10^{-17}$ erg~s$^{-1}$cm$^{-2}$arcsec$^{-2}$.
Therefore, if we have such large ELANe in the obtained quasar-nebulae, it is expected to see Ly$\alpha$ emission extending beyond $\sim20$ arcsec in their residual Ly$\alpha$ images (figure~\ref{fig8} and \ref{fig1a}).
However, apparently none of them represent such extensive Ly$\alpha$ nebulae significantly beyond 100 pkpc above the surface brightness limit of $\sim10^{-17}$ erg~s$^{-1}$cm$^{-2}$arcsec$^{-2}$, despite the large sample of luminous quasars ($M_i<-27$) amounting to 29 and 731 objects.
The most extended nebula in the current sample is {\tt object\_id=37489365072502768} in the DUD layer (the second line in figure~\ref{fig8}), which has the maximum extent of 110 pkpc above $\sim7\times10^{-18}$ erg~s$^{-1}$cm$^{-2}$arcsec$^{-2}$ according to the tentative flux estimation (eq.~\ref{eq3}).
This may suggest that the largest class of ELANe as ever discovered could be extremely rare ($\ll1$ percent) than previously thought \citep{ArrigoniBattaia2019}.
The upcoming huge imaging data by the LSST on the Vera C. Rubin Observatory over 18000 deg$^2$ \citep{Ivezic2019}, 13 times wider than the survey area of the HSC-SSP (1400 deg$^2$), may be necessary to establish statistical ELAN samples with the broad-band selection.
Particularly, the LSST will also conduct imaging with $u$-band, which has $\sim2$ times narrower filter width than the $g$-band, and hence more sensitive and useful to detect the Ly$\alpha$ excess at $z\sim2$.


\section{Conclusions}
\label{s6}

We applied the broad-band colour selection to all available SDSS-IV/eBOSS quasars at $z=$ 2.34--3.00 from \citep{Lyke2020} in the HSC-SSP PDR3 field \citep{Aihara2022}, amounting to the total of 8683 sources, to search for extended Ly$\alpha$ emission associated with them.
Our broad-band selection could extract only bright Ly$\alpha$ emission approximately down to $1\times10^{-17}$ erg~s$^{-1}$cm$^{-2}$arcsec$^{-2}$. 
However, we had a huge advantage of adopting the very wide-field imaging data covering 35 and 890 deg$^2$ in the HSC-SSP DUD and Wide layers, allowing us to estimate the extent of Ly$\alpha$ emission for two orders of magnitude more sources than any existing narrow-band imaging and IFU spectroscopic surveys.

Consequently, we measured Ly$\alpha$ areas for all 8683 targets, and then selected 7 and 32 quasars with extended Ly$\alpha$ emission larger than 40 arcsec$^2$ as quasar-nebulae in the DUD and Wide layers, respectively.
Although some of them may be potential ELAN candidates, none of them have remarkably large Ly$\alpha$ nebulae as seen in the Slug and MAMMOTH-1 nebulae \citep{Cantalupo2014,Cai2017}, suggesting that such a largest class of ELANe could be extremely rare in the universe ($\ll1$ percent).
The source catalogue of our quasar targets and their Ly$\alpha$ properties derived through this paper are available as online material, and their processed images (i.e., $gri$ colour, $G$, $RI_z$, and $G-RI_z$ images) also appear in figure~\ref{fig8} and \ref{fig1a}.
Based on the obtained effective Ly$\alpha$ areas of quasars, we detected higher fractions of relatively large Ly$\alpha$ nebulae in more luminous and radio-loud quasars by reference to the public catalogues \citep{Becker1995,White1997,Lyke2020}.
On the other hand, overall there is no statistical difference of effective Ly$\alpha$ areas between radio-loud and radio-quiet quasars.
These obtained trends of the extent of Ly$\alpha$ emission towards luminosity and radio-loudness are broadly consistent with previous findings with deep IFU observations for $<100$ quasars at $z>2$ \citep{ArrigoniBattaia2019,Cai2019,Farina2019,Mackenzie2021}.
Forthcoming paper II will address their Ly$\alpha$ morphologies and environmental dependence, including additional $u$-band selected quasar-nebulae at $z\sim2$, based on a deep multi-band photometric catalogue in the DUD layer.

Catalogue users can easily access the reduced coadd images of our sample on the HSC-SSP data release site (see the data availability section).
The source catalogue (table~\ref{tab2}) also includes the spatial properties of Ly$\alpha$ emissions for the non quasar-nebulae, allowing users to crosscheck their samples with all the quasars adopted in this study.
We note that {\tt SB\_ann} is the preliminary estimate of the surface Ly$\alpha$ brightness with large uncertainties.
Users are advised to adopt this value at their own risk.
\emph{It will be very helpful if catalogue users share success rates of their follow-up observations, which will improve the selection method for future work.}
We plan to extend the sample to available spec-$z$ sources, including non-quasar objects at lower and higher redshifts by using more broad-band data.
Such a series of efforts will help us scrutinise the redshift evolution and environmental dependence of large Ly$\alpha$ nebulae, and greatly push this field forward with the LSST on the Vera C. Rubin Observatory in the near future \citep{Ivezic2019}.

\section*{Acknowledgements}

This research is based on data collected at Subaru Telescope, which is operated by the National Astronomical Observatory of Japan.
We are honoured and grateful for the opportunity of observing the Universe from Maunakea, which has the cultural, historical and natural significance in Hawaii.

The Hyper Suprime-Cam (HSC) collaboration includes the astronomical communities of Japan and Taiwan, and Princeton University. The HSC instrumentation and software were developed by the National Astronomical Observatory of Japan (NAOJ), the Kavli Institute for the Physics and Mathematics of the Universe (Kavli IPMU), the University of Tokyo, the High Energy Accelerator Research Organization (KEK), the Academia Sinica Institute for Astronomy and Astrophysics in Taiwan (ASIAA), and Princeton University. Funding was contributed by the FIRST program from Japanese Cabinet Office, the Ministry of Education, Culture, Sports, Science and Technology (MEXT), the Japan Society for the Promotion of Science (JSPS), Japan Science and Technology Agency (JST), the Toray Science Foundation, NAOJ, Kavli IPMU, KEK, ASIAA, and Princeton University. 
This paper makes use of software developed for the Large Synoptic Survey Telescope. We thank the LSST Project for making their code available as free software at \url{http://dm.lsst.org}.

The Pan-STARRS1 Surveys (PS1) have been made possible through contributions of the Institute for Astronomy, the University of Hawaii, the Pan-STARRS Project Office, the Max-Planck Society and its participating institutes, the Max Planck Institute for Astronomy, Heidelberg and the Max Planck Institute for Extraterrestrial Physics, Garching, The Johns Hopkins University, Durham University, the University of Edinburgh, Queen’s University Belfast, the Harvard-Smithsonian Center for Astrophysics, the Las Cumbres Observatory Global Telescope Network Incorporated, the National Central University of Taiwan, the Space Telescope Science Institute, the National Aeronautics and Space Administration under Grant No. NNX08AR22G issued through the Planetary Science Division of the NASA Science Mission Directorate, the National Science Foundation under Grant No. AST-1238877, the University of Maryland, and Eotvos Lorand University (ELTE) and the Los Alamos National Laboratory.

Funding for the Sloan Digital Sky Survey IV has been provided by the Alfred P. Sloan Foundation, the U.S. Department of Energy Office of Science, and the Participating Institutions. SDSS-IV acknowledges support and resources from the Center for High Performance Computing  at the University of Utah. The SDSS website is \url{www.sdss.org}.

SDSS-IV is managed by the Astrophysical Research Consortium for the Participating Institutions of the SDSS Collaboration including the Brazilian Participation Group, the Carnegie Institution for Science, Carnegie Mellon University, Center for Astrophysics | Harvard \& Smithsonian, the Chilean Participation Group, the French Participation Group, Instituto de Astrof\'isica de Canarias, The Johns Hopkins University, Kavli Institute for the Physics and Mathematics of the Universe (IPMU) / University of Tokyo, the Korean Participation Group, Lawrence Berkeley National Laboratory, Leibniz Institut f\"ur Astrophysik Potsdam (AIP),  Max-Planck-Institut f\"ur Astronomie (MPIA Heidelberg), Max-Planck-Institut f\"ur Astrophysik (MPA Garching), Max-Planck-Institut f\"ur Extraterrestrische Physik (MPE), National Astronomical Observatories of China, New Mexico State University, New York University, University of Notre Dame, Observat\'ario Nacional / MCTI, The Ohio State University, Pennsylvania State University, Shanghai Astronomical Observatory, United Kingdom Participation Group, Universidad Nacional Aut\'onoma de M\'exico, University of Arizona, University of Colorado Boulder, University of Oxford, University of Portsmouth, University of Utah, University of Virginia, University of Washington, University of Wisconsin, Vanderbilt University, and Yale University.

GAMA is a joint European-Australasian project based around a spectroscopic campaign using the Anglo-Australian Telescope. The GAMA input catalogue is based on data taken from the Sloan Digital Sky Survey and the UKIRT Infrared Deep Sky Survey. Complementary imaging of the GAMA regions is being obtained by a number of independent survey programmes including GALEX MIS, VST KiDS, VISTA VIKING, WISE, Herschel-ATLAS, GMRT and ASKAP providing UV to radio coverage. GAMA is funded by the STFC (UK), the ARC (Australia), the AAO, and the participating institutions. The GAMA website is \url{www.gama-survey.org}.
This work is in part based on observations taken by the 3D-HST Treasury Program (GO 12177 and 12328) with the NASA/ESA HST, which is operated by the Association of Universities for Research in Astronomy, Inc., under NASA contract NAS5-26555.
Funding for PRIMUS is provided by NSF (AST-0607701, AST-0908246, AST-0908442, AST-0908354) and NASA (Spitzer-1356708, 08-ADP08-0019, NNX09AC95G).
This research uses data from the VIMOS VLT Deep Survey, obtained from the VVDS database operated by Cesam, Laboratoire d'Astrophysique de Marseille, France.

We thank anonymous referee for helpful feedback.
We would like to thank Editage (\url{www.editage.com}) for English language editing.
This work made extensive use of the following tools, {\tt NumPy} \citep{Harris2020}, {\tt Matplotlib} \citep{Hunter2007}, the Tool for OPerations on Catalogues And Tables, {\tt TOPCAT} \citep{Taylor2005}, a community-developed core Python package for Astronomy, {\tt Astopy} \citep{AstropyCollaboration2013}, and Python Data Analysis Library {\tt pandas} \citep{Reback2021}.

\section*{Data Availability}

The data underlying this article are available on the public data release site of Hyper Suprime-Cam Subaru Strategic Program (\url{https://hsc.mtk.nao.ac.jp/ssp/data-release/}).
The source catalogue and processed images are also accessible as online material.
When one wants to obtain coadd imaging data for specific sources, the author recommends using the user-friendly cutout tool (\url{https://hsc-release.mtk.nao.ac.jp/das_cutout/pdr3/}) or the interactive sky viewer {\tt hscMap} (\url{https://hsc-release.mtk.nao.ac.jp/hscMap-pdr3/app}).
For any additional information, including the risk catalogue for $z=$ 3.0--3.5 quasars, please contact the author.



\bibliographystyle{mnras}
\bibliography{rs22a} 



\appendix

\section{Additional figures}

\begin{figure*}
\centering
\includegraphics[width=14cm]{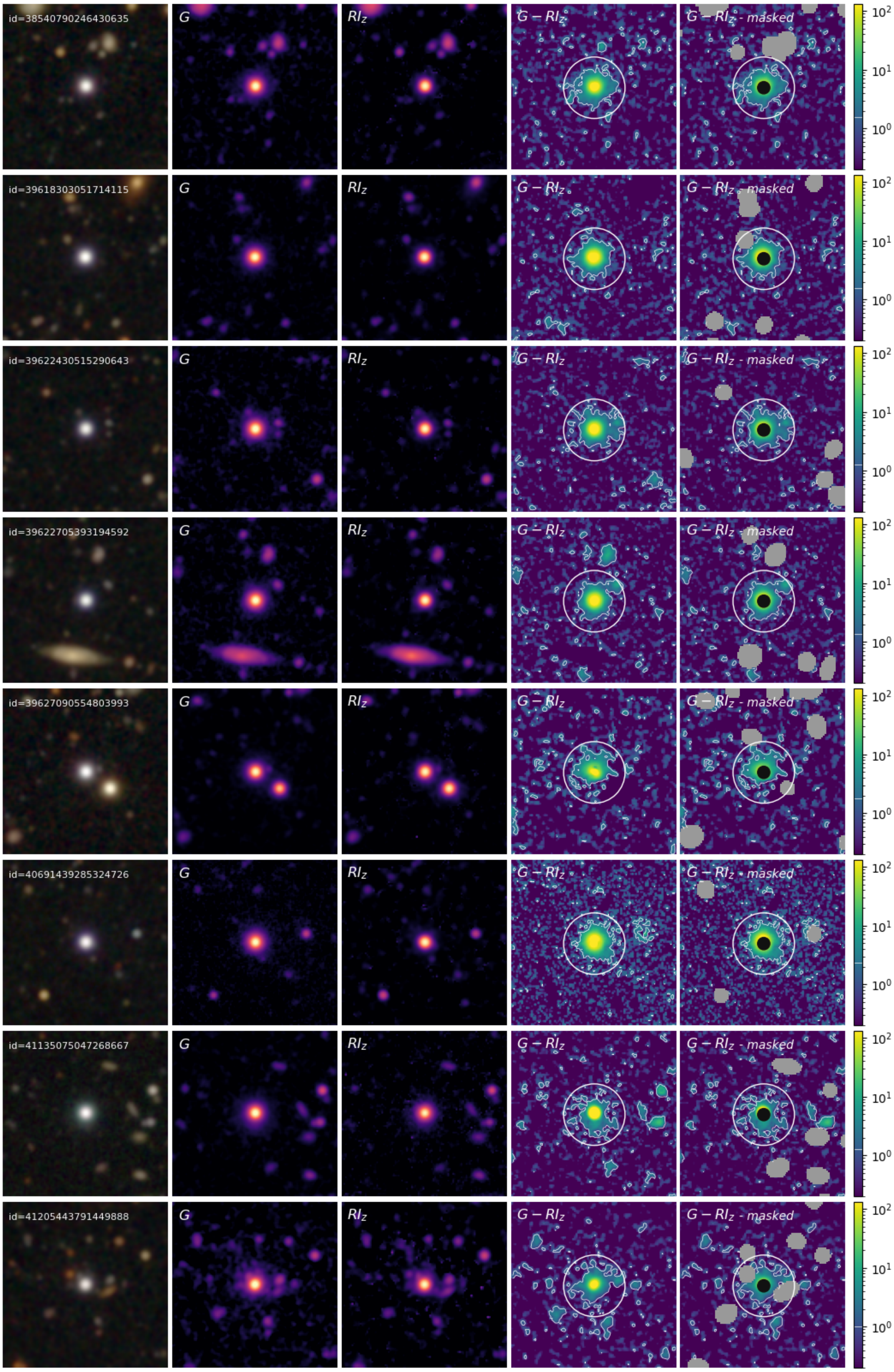}
\caption{
Same as figure~\ref{fig8}, but in the HSC-SSP PDR3 Wide layer. 
}
\label{fig1a}
\end{figure*}

\begin{figure*}
\begin{center}
\includegraphics[width=14cm]{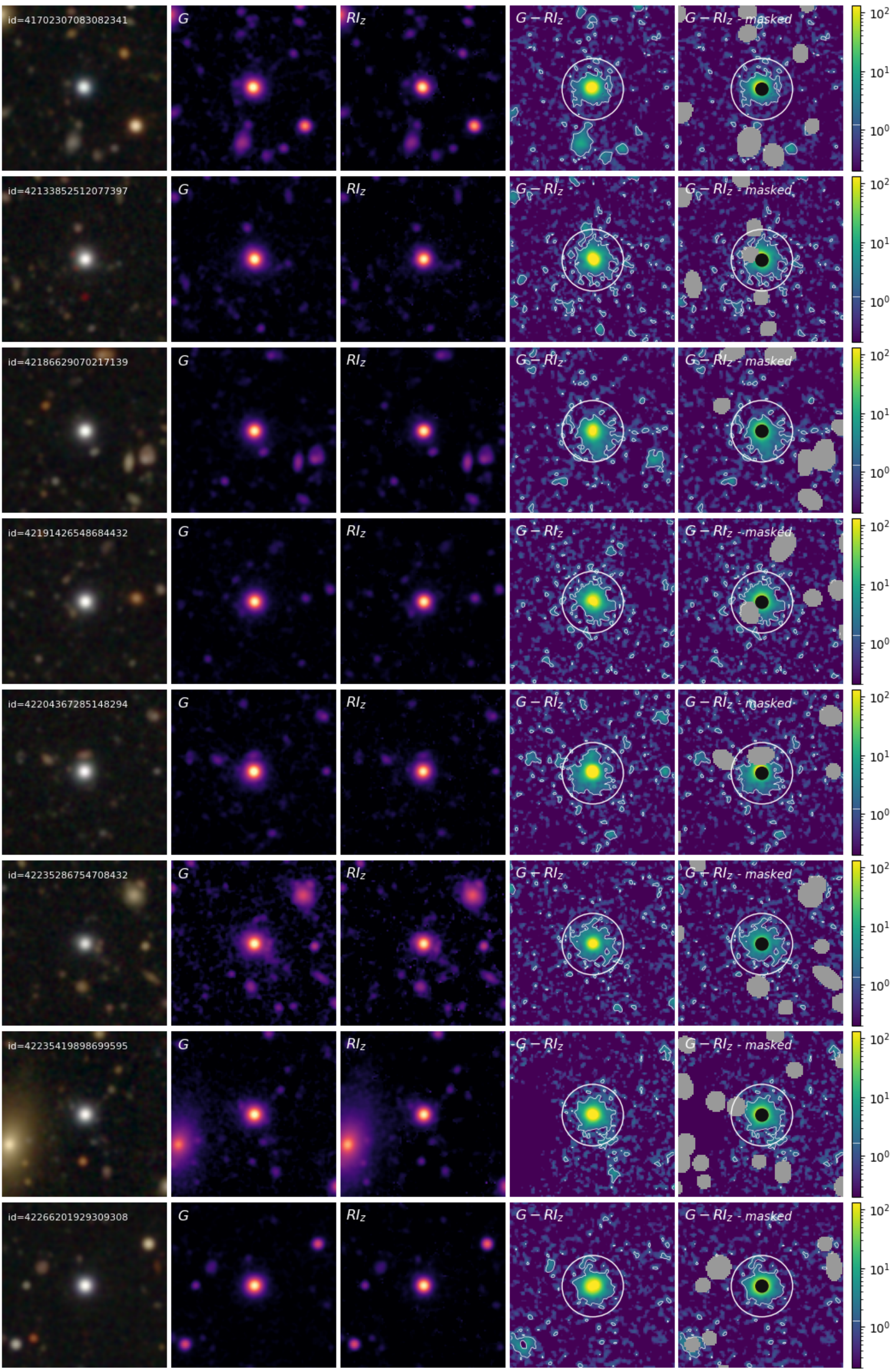}
\end{center}
\contcaption{}
\end{figure*}

\begin{figure*}
\begin{center}
\includegraphics[width=14cm]{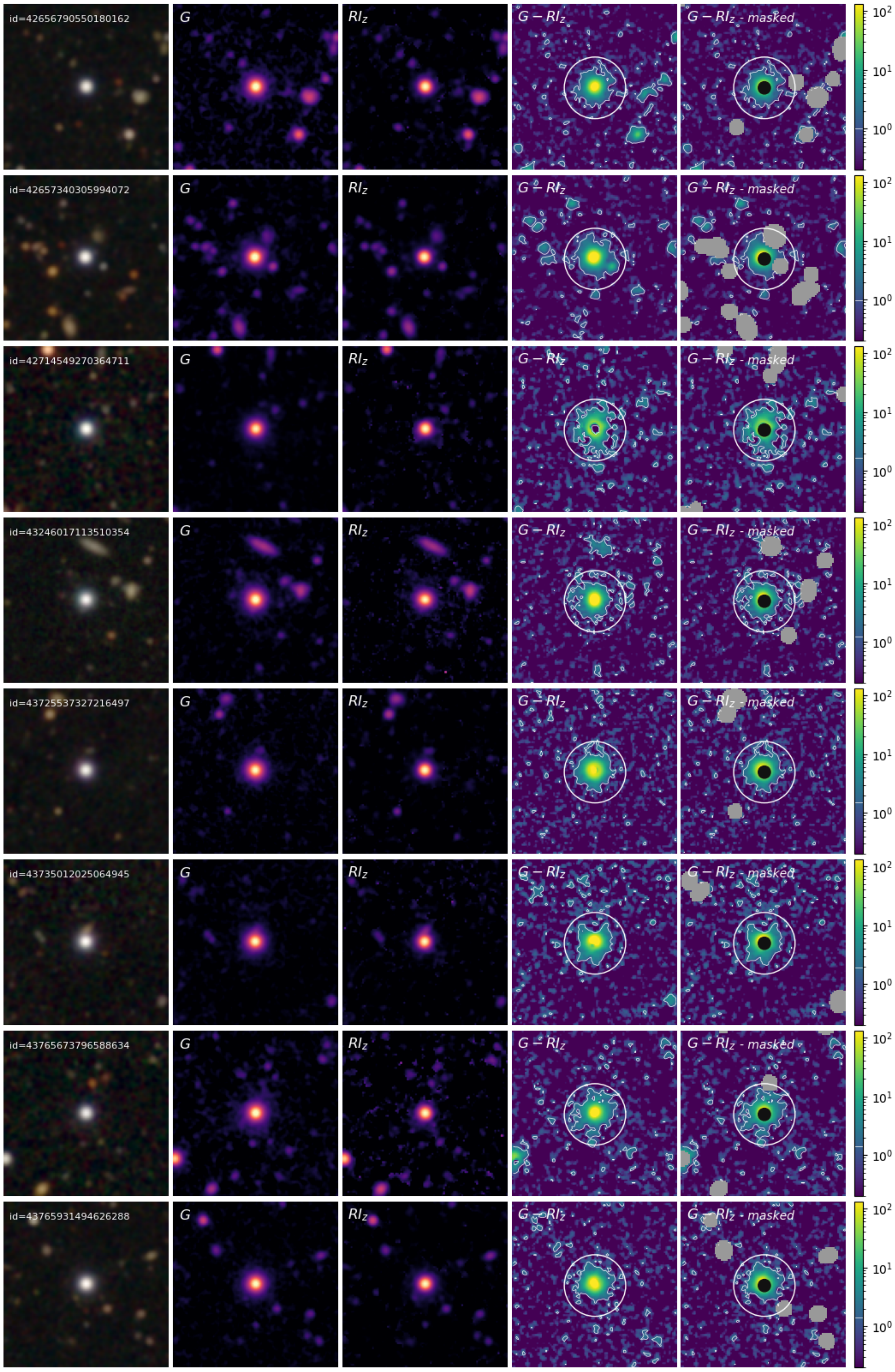}
\end{center}
\contcaption{}
\end{figure*}

\begin{figure*}
\begin{center}
\includegraphics[width=14cm]{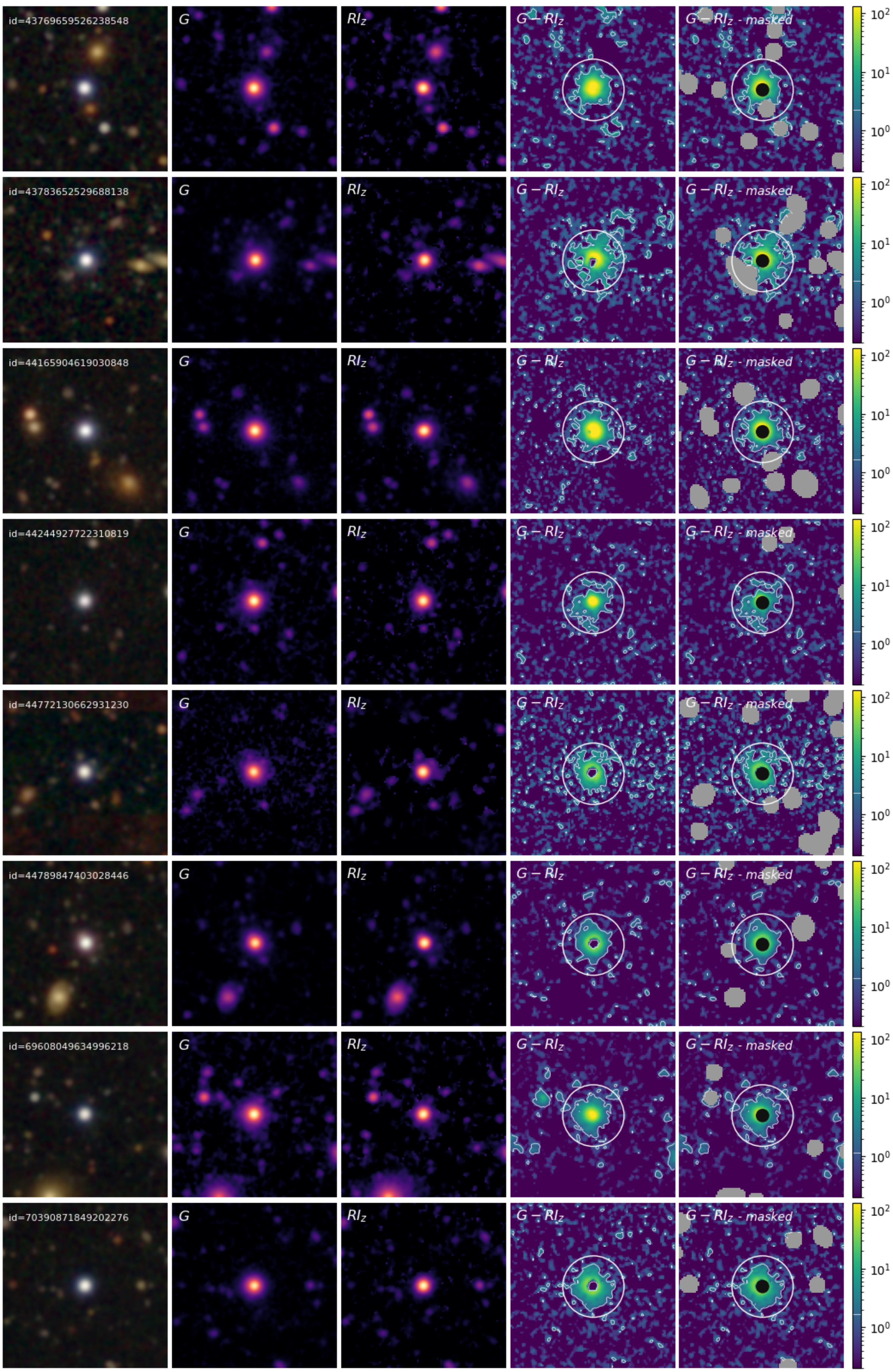}
\end{center}
\contcaption{}
\end{figure*}


\bsp	
\label{lastpage}
\end{document}